\documentclass[prl,aps,reprint,twocolumn,superscriptaddress,footnoteinbib,showpacs,floatfix]{revtex4-1}
\usepackage{textcomp}
\usepackage{amssymb,amsmath,MnSymbol,ifsym,latexsym,bm}
\usepackage{wrapfig,subfigure}
\usepackage[Euler]{upgreek}
\usepackage{longtable, supertabular}

\bibpunct[, ]{[}{]}{;}{n}{,}{,}

\usepackage[pdftex]{graphics}
\usepackage[pdftex,final]{graphicx}

\newcommand{\etas}{\upeta_\mathrm{s}}

\newcommand{\eeta}{\overline{\upeta}}
\newcommand{\eetap}{\overline{\upeta}_\mathrm{p}}

\newcommand{\kB}{k_{\textsc{b}}}
\newcommand{\kBT}{k_{\textsc{b}} \,T}

\newcommand{\edot}{\dot{\varepsilon}}

\newcommand{\Wi}{\mathrm{W\!i}}
\newcommand{\Wics}{\Wi_{\mathrm{c-s}}}
\newcommand{\Wisc}{\Wi_{\mathrm{s-c}}}

\newcommand{\NK}{N_\textsc{k}}

\newcommand{\bK}{b_\textsc{k}}
\newcommand{\hK}{h^\ast_\textsc{k}}

\newcommand{\lamo}{\lambda_0}

\newcommand{\bQ}{\bf Q}

\newcommand{\Mtens}{\langle \widetilde{\bQ} \widetilde{\bQ} \rangle}

\newcommand{\zetao}{\zeta_0}
\newcommand{\zetas}{\zeta_\mathrm{s}}
\newcommand{\zetaZ}{\zeta_\textsc{z}}
\newcommand{\zetaR}{\zeta_\textsc{r}}

\newcommand{\zetat}{\zeta_\mathrm{t}}
\newcommand{\zetac}{\zeta_\mathrm{c}}

\newcommand{\cstar}{c^\ast}
\newcommand{\cdags}{c^\dagger_s}
\newcommand{\cdagc}{c^\dagger_c}

\newcommand{\phit}{\phi_\mathrm{t}}

\newcommand{\xit}{\xi_\mathrm{t}}
\newcommand{\Nt}{N_\mathrm{t}}
\newcommand{\xic}{\xi_\mathrm{c}}

\newcommand{\Nc}{N_\mathrm{c}}

\begin{document}

\title{Effect of stretching-induced changes in hydrodynamic screening on coil--stretch hysteresis of unentangled polymer solutions}
\author{Ranganathan Prabhakar}
\affiliation{Department of Mechanical \& Aerospace Engineering, Monash University, Clayton,  VIC 3800, AUSTRALIA\\
email: prabhakar.ranganathan@monash.edu}
\author{Chandi Sasmal}
\affiliation{Department of Chemical Engineering, Monash University, Clayton,  VIC 3800, AUSTRALIA}
\author{Duc At  Nguyen}
\affiliation{Department of Chemical Engineering, Monash University, Clayton,  VIC 3800, AUSTRALIA}
\author{Tam Sridhar}
\affiliation{Department of Chemical Engineering, Monash University, Clayton,  VIC 3800, AUSTRALIA}
\author{J. Ravi Prakash}
\affiliation{Department of Chemical Engineering, Monash University, Clayton,  VIC 3800, AUSTRALIA}

\date{\today}

\begin{abstract}
Extensional rheometry and Brownian Dynamics simulations of flexible polymer solutions  confirm predictions based on blob concepts that coil--stretch hysteresis in extensional flows increases with concentration, reaching a maximum at the critical overlap concentration $\cstar$ before progressively vanishing in the semidilute regime. These observations demonstrate that chain stretching strengthens intermolecular hydrodynamic screening in dilute solutions, but weakens it in semidilute solutions. Flow can thus strongly modify the concentration dependence of viscoelastic properties of polymer solutions.
 \end{abstract}

\maketitle

The molecular mechanisms underlying the dynamics of flexible polymers in solution have long held the fascination of physicists \citep{rubinsteincolby}. Intramolecular hydrodynamic interactions (HI)  play a central role in  determining mechanical properties of dilute polymer solutions \citep{Zimm}. 
The progressive emergence of screening of hydrodynamic (or of excluded-volume) interactions with polymer concentration in semidilute polymer solutions remains relatively unexplored in comparison with phenomena such as entanglements or reptation. Most of our current understanding comes from studies under isotropic conditions close to equilibrium, but how does stretching in strong flows affect screening? The answer to this question is essential to understand how  macroscopic properties depend on polymer concentration in semidilute solutions. This is also significant for applications such as turbulent drag reduction \citep{White2008}, inkjet printing \citep{Sirringhaus2000}, and electrospinning \citep{Wang2006}, where it is necessary to optimize polymer concentration to achieve good performance.

We use 
the phenomenon of coil--stretch hysteresis in extensional flows  to gain insight into screening of HI in strong flows of polymer solutions. Steady state in extensional flow is primarily the result of a balance between internal resistance of polymer molecules to stretching  and the frictional drag force exerted on molecules by the flowing solvent. The relative strength of an extensional flow is expressed in terms of the Weissenberg number,  $\Wi \equiv \edot \, \lamo$, where $\edot$ is the strain rate of the imposed flow and $\lamo$ is the time-scale of the slowest relaxation mode of a polymer molecule in a quiescent solution.  The  friction coefficient of an isolated polymer molecule changes as it is stretched because intramolecular HI and the shielding effect it provides weaken with stretching, exposing segments to solvent flow.   \citet{degennes}, \citet{hinch}  and  \citet{tanner} showed that a  consequence of such conformation-dependent friction is that, at any $\Wi$ within a window  $\Wisc < \Wi < \Wics$ in extensional flows of dilute polymer solutions, there are two stable states for polymer conformations -- a coiled state and a stretched state. Here, $\Wics$ and $\Wisc$ are the critical values for the coil-to-stretch and stretch-to-coil transitions. Outside this window, only a single state is stable. 
These predictions have been extensively verified in the dilute regime through observations of strong hysteresis in chain conformations in single-molecule simulations  \citep{schroeder} and experiments on single DNA molecules \citep{smithchu, schroeder_science, schroeder, Shaqfeh2005}, and in rheological properties measured with the filament-stretching extensional rheometer (FiSER)  \citep{sridharPRL, larsonreview}. 

\begin{figure}[h]
\centerline{\resizebox{8.3cm}{!}{\includegraphics{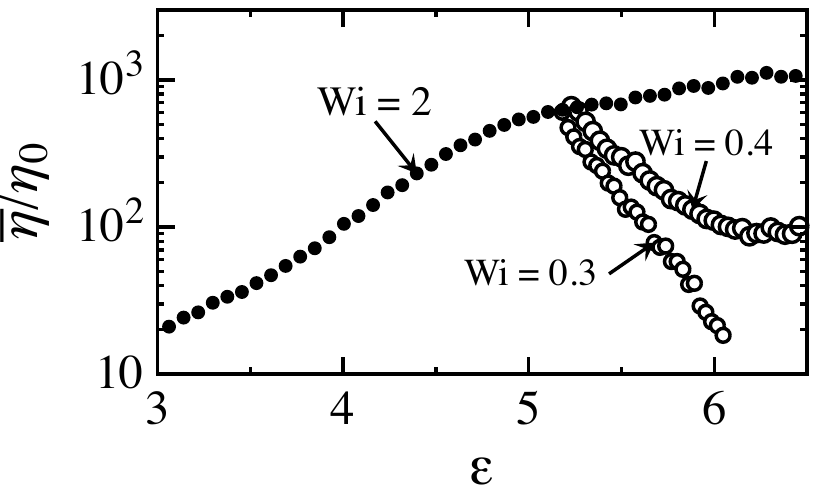}}}
\caption{\label{f:quench} Transient evolution of $\eeta$ normalized by the zero-shear rate solution viscosity $\eta_0$ to steady state at $\Wi = 2$ (filled circles)), and following strain-rate quenches (open circles) to lower values of $\Wi$: the stretch-to-coil transition is identified by noting the quench $\Wi$ below which $\eeta$ relaxes continuously without reaching a steady-state within the maximum Hencky strain achievable of about 7. }
\end{figure}

\begin{figure*}[h]
\centerline{\resizebox{17cm}{!}{\includegraphics{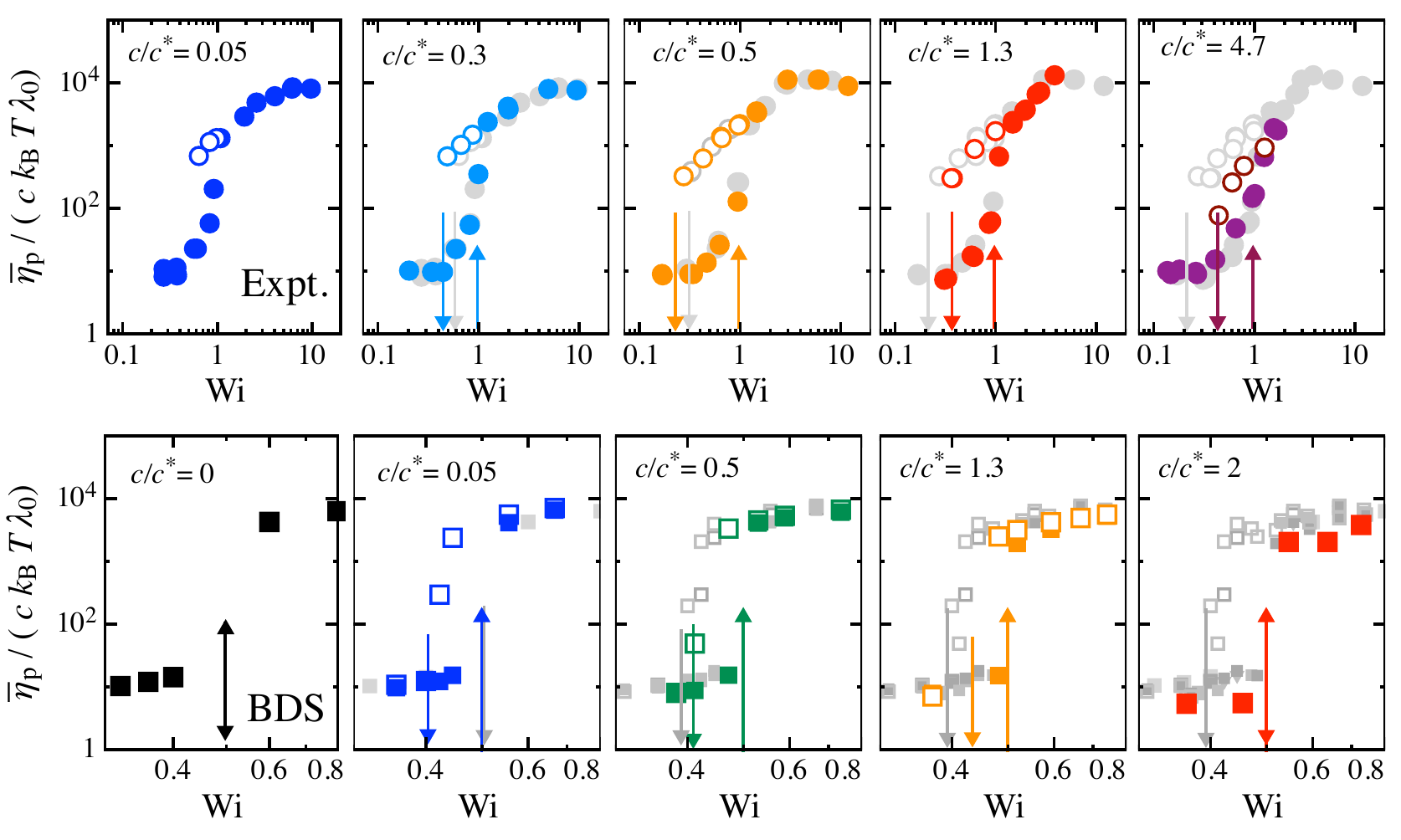}}}
\caption{\label{f:exptbdscsh} Concentration dependence of coil--stretch hysteresis in  $\eetap$, the polymer contribution to extensional viscosity, in (a) FiSER experiments (top panel)  and (b) multi-chain Brownian Dynamics simulations (bottom panel). Coiled and stretched states are represented by filled and open symbols respectively. Concentration increases from left to right.  Coloured upward and downward arrows indicate $\Wics$ and $\Wisc$, respectively. Gray symbols are data from sub-plots at lower concentrations while gray arrows indicate the smallest $\Wisc$ from those sub-plots.  The number of Kuhn segments $\NK = L^2/R_0^2 \sim 1300$ in the experiments and simulations  \citep{suppmatl}.}
\end{figure*}

The question is: how does coil--stretch hysteresis change with polymer concentration, $c$? We performed uniaxial extensional flow experiments with the FiSER using solutions of high-molecular weight polystyrene in a solvent consisting of a mixture of oligomeric styrene and dioctylphthalate \citep{sridharPRL, gupta, suppmatl}. In these experiments, polymer solution samples sandwiched between a pair of end-plates were stretched by moving the plates rapidly to create  slender filaments. The stretching rate was controlled such that a predetermined constant $\edot$ was obtained at the necking plane. The polymer contribution to the viscoelastic stress at the neck was calculated from the force measured on the end-plate \citep{Tirtaatmadja1993, mckinleysridhar}. Figure~\ref{f:quench} plots the transient evolution of the transient extensional viscosity $\eeta$ with Hencky strain $\varepsilon$ for a typical sample. Two sets of experiments were conducted. In the first, the viscosity was allowed to saturate to give the steady-state viscosity at the $\Wi$ corresponding to the strain-rate at the neck. The relaxation times $\lamo$ for all samples were obtained by small-amplitude oscillatory shear rheometry \citep{suppmatl}. In the second set of experiments, after stretching initially at $\Wi = 2 \,(> \Wics \approx 1/2)$, the neck strain-rate was quenched at a Hencky strain of $5$ to a new lower value. Steady-state viscosities were obtained or a range of $\Wi$ values. It was possible to detect an abrupt transition $\Wisc$ below which the polymer stress after quench continually relaxed without reaching a plateau (Fig.~\ref{f:quench}). Figure~\ref{f:exptbdscsh} (a) plots (as solid symbols) the steady-state polymer contribution to the extensional viscosity $\eetap = \eeta - 3 \etas$ (where $\etas$ is the shear viscosity of the Newtonian solvent) against the corresponding $\Wi$ at the steady-state. A  coil-to-stretch transition is observed at $\Wics \sim 1/2$. We find that the size of the coil--stretch hysteresis window does not decrease monotonically with concentration when going from dilute to concentrated solutions.  When $c/\cstar < 1$, the stretch-to-coil transition at a $\Wisc$ is first observed to \textit{decrease} with increasing concentration. The hysteresis loop thus widens and only begins to shrink in size when $c \gtr \cstar$, where $\cstar$ is the concentration at which coils overlap at equilibrium. 

We have also observed a similar non-monotonic variation of the hysteresis window in multi-chain Brownian Dynamics (BD) simulations of polymer solutions (Fig.~\ref{f:exptbdscsh} (b)). A single polymer molecule in these simulations was represented as a chain of beads connected by  FENE springs. An optimized Ewald algorithm was used to calculate pair-wise HI between beads of all chains in a simulation box. Periodic Kraynik-Reinhelt boundary conditions were used to simulate large strains in planar extensional flows \citep{jainflow}.   Simulation parameters were chosen to resemble the experimental system. Ensembles with chains initially in their equilibrium configurations and with highly stretched chains were separately simulated to obtain the coiled and stretched-state branches of the hysteresis windows, respectively \citep{suppmatl}. Excluded volume interactions were neglected. 

\begin{figure}[h]
\centerline{\resizebox{8.3cm}{!}{\includegraphics{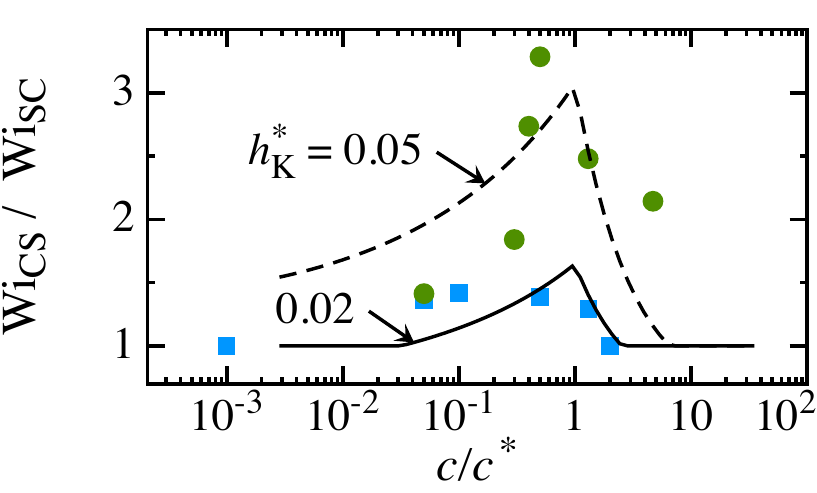}}}
\caption{\label{f:wiscratio}  Concentration dependence of the width of the hysteresis window observed in FiSER experiments (green circles) and BD simulations (blue squares): the continuous and broken curves are predictions with the blob model \citep{Prabhakar2016} for the hydrodynamic interaction parameter \citep{suppmatl} $\hK  = 0.02$ and 0.05, respectively; all data are for the number of Kuhn segments \citep{suppmatl}, $\NK \sim 1300$.}
\end{figure}

The increase in the hysteresis size $\Wics/\Wisc$  in the dilute regime before vanishing in semidilute solutions (Fig.~\ref{f:wiscratio}) is unexpected. The ratio $\Wics/\Wisc$ quantifying the size of the hysteresis is known to be proportional to the ratio of $\zetas$, the average friction coefficient of a chain stretched close to its contour length $L$, to $\zetao$, the coefficient for an isotropic coil of radius $R_0$ at equilibrium \citep{degennes, schroeder_science,schroeder,hsiehlarsoncsh}. Coil--stretch hysteresis can only be observed if $\zetas$ is significantly larger than $\zetao$. In the absence of significant intermolecular interactions, $\zetas$ is larger than $\zetao$ because the shielding due to intramolecular HI of segments in the interior of polymer coils from the solvent velocity field weakens when chains are stretched in flow.  In a concentrated solution on the other hand where HI is highly screened, chains are expected to be freely-draining. The  friction coefficient of  such Rouse chains is the sum of contributions from every segment irrespective of chain conformation, and $\zetao \sim \zetas  \sim \etas \, L$:  no hysteresis is therefore expected in concentrated solutions. 

It is possible that the peak in the ratio $\Wics/\Wisc \sim \zetas/ \zetao$ is observed because $\zetas$ and $ \zetao$ approach the Rouse limit at different rates with increasing concentration \citep{Prabhakar2016}. The concentration dependence of $\zetao$ is understood through the scaling theory proposed by \citet{degennes} which argues that, when isotropic coils interpenetrate at concentrations above $\cstar \sim R_0^{-3}$, the hydrodynamic screening length $\xic$ is estimated as the scale at which the mean intramolecular segmental density in a typical chain is equal to the mean segmental density of the whole solution \citep{rubinsteincolby}.  Segments separated by distances larger than $\xic$ only experience Rouse-like correlations solely due to backbone connectivity. This scaling theory suggests that $\zetao$ is nearly constant at the value $\zetaZ$ calculated by Zimm hydrodynamics for isolated coils until coils begin to overlap at $\cstar$.  Thereafter, each chain can therefore be pictured as a Rouse chain of $\Nc$ ``correlation blobs" of size $\xic$, where for polymer coils at the theta state, $\xic \sim R_0 \, (c/\cstar)^{-1}$ and $\Nc \sim (c/\cstar)^{2}$ \citep{Prabhakar2016}.  Since residual intramolecular HI persists within each blob, the friction coefficient of a single correlation blob $\zetac \sim \zetaZ \, \xic/ R_0 \sim \etas \xic$. Consequently, the sum of the friction coefficients is $\zetao \sim \zetac \Nc \sim \zetaZ \, (c/\cstar)$, when $c/\cstar > 1$.


\citet{pincus1976} suggested that stretching brings into play another characteristic length scale $\xit$ which is the result of the competition between  random thermal forces and the tension in the chain. In a chain that is stretched to a length of $R \gg R_0$, the influence of the applied tension in creating anisotropy in segmental orientation becomes apparent only across scales larger than $\xit \sim R_0^2/R$ under theta conditions. Stretched molecules can thus be regarded as chains of $\Nt \sim (R/R_0)^2$ ``tension blobs" each of size $\xit$. Although conformational fluctuations in the direction transverse to the extensional axis are not strongly suppressed when $\Wi \lesssim 1$,  the friction coefficient of stretched chains in dilute solutions is well approximated  by that of a linear array of tension blobs or a ``blob pole" of length $R$ and diameter $\xit$  \citep{Prabhakar2016}. Thus, if $\zetat \sim \etas \xit$ is the friction coefficient of a single tension blob, then from slender body hydrodynamics \citep{batchelor1}, $\zetas \sim \zetat \Nt/ \ln (2 \Nt) \sim \etas R/ \ln (2 \Nt)$.

It is well known that stretched chains interact significantly with each other hydrodynamically at concentrations such that $c R^3 \sim 1$, well before they physically overlap and interpenetrate  \citep{clasenetal}. The boundary between the dilute and semidilute regime is thus given by $c/\cstar \sim ( R/ R_0)^{-1/3}$. Transverse fluctuations of size $R_0$ in stretched chains can further cause them to overlap  when $c/\cstar \sim R_0/ R$. Screening of HI in such \textit{partially} stretched chains  overlapping has not been widely studied.  At incipient transverse overlap, the correlation blobs that emerge are anisotropic since the size of correlation blobs is comparable to that of the whole stretched chain and much larger than a tension blob.  Scaling analysis \citep{Prabhakar2016} suggests that despite transverse chain overlaps, as long as correlation blobs are significantly larger than tension blobs, the solution continues to behave as a suspension of tension-blob poles with a screening length $\chi \sim \xit\, [(1/\phit)  \ln (1/\phit)]^{1/2} $, where $\phit = c R \xit^2$ is the volume fraction of tension-blob poles. The friction coefficient then is $\zetas \sim \zetat \Nt/ \ln (2 \chi/ \xit )$.

Correlation blobs become smaller and less anisotropic with increasing concentration, until they become comparable in size to tension blobs.  Beyond that, correlation blobs are smaller than tension blobs and are hence isotropic. Their size and number are insensitive to stretching, and their values are identical to those calculated for equilibrium coils at the same  $c/\cstar$. In this regime, the screening length is entirely set by the isotropic correlation blob size $\xic$.  The crossover from the weak screening between tension-blob poles to the stronger screening between correlation blobs occurs when $\xic \sim \xit$ at concentrations $c/\cstar \gtrsim R/R_0$ \citep{suppmatl}. At this boundary, $\zetas \sim \zetat \Nt \sim \zetac \Nc$: in other words, increasing concentration across the regime of weakly-screened HI just gets rid of the logarithmic correction to the Rouse-like drag coefficient of a stretched chain in a dilute solution. Beyond this boundary,  molecules are Rouse chains of correlation blobs and friction is conformation independent \textit{i.e.} $\zetas \sim \zetao$. 

As a result of the weak-screening of intramolecular HI in stretched chains, $\zetas$ can increase logarithmically with concentration even when $c/\cstar \ll 1$, where $\zetao$ is nearly constant \citep{suppmatl}. Under such conditions, the ratio $\zetas/\zetao$ increases. When $c/\cstar >1$ on the other hand, stretched chains can still be  weakly screened, whereas coiled chains at the same concentration are strongly screened. Since $\zetao$ increases linearly whereas $\zetas$ changes more slowly in this case, $\zetas/\zetao$ decreases with $c$ until even the stretched state is strongly screened. Thus, $\zetas/\zetao$, and therefore coil--stretch hysteresis is maximal at $c/\cstar \sim 1$.  Figure~\ref{f:wiscratio} shows good qualitative agreement of the $\zetas/\zetao$ ratio predicted by the scaling model of \citet{Prabhakar2016}, with the hysteresis widths $\Wics/\Wisc$ observed in the experiments and simulations. This agreement appears to confirm the existence of a transition from weak to strong hydrodynamic screening in stretched chains. In a nominally dilute solution with $c/\cstar \ll 1$, the stronger hydrodynamic interaction between chains when stretched means that macroscopic properties in a dilute solution can exhibit greater sensitivity to concentration in strong flows than at equilibrium \textit{i.e.} dilute solutions can ``self-concentrate" \citep{clasenetal}. The results presented above imply that unentangled semidilute solutions can conversely ``self-dilute'', with the concentration dependence weakening in strong flows. 

The data in Fig.\ref{f:exptbdscsh} and \ref{f:wiscratio} show that in either the experiments or simulations, increasing concentration can cause a hysteresis to emerge when no hysteresis exists in the dilute limit. The friction coefficient of isolated chains in these systems appears not to change significantly even though chains stretch by more than an order of magnitude in extensional flow.  The blob model suggests this may be due to the small values of the hydrodynamic radius of the Kuhn segment in these systems, quantified by the dimensionless parameter $\hK$. This parameter has little influence on the friction of isolated coils: the prefactor to the scaling result $\zetaZ \sim \etas R_0$ has a universal value that is insensitive to $\hK$ when $\NK$ is large. In contrast, as the tension blob in a dilute solution shrinks with increasing polymer stretch and becomes comparable in size to a single Kuhn segment, its friction coefficient --- and hence the prefactor to the scaling result for $\zetas$ --- is directly proportional to $\hK$ \citep{suppmatl}. If $\hK$ is small enough, the prefactor can decrease with chain stretching to compensate the contribution to friction from the change in length. At concentrations comparable to $\cstar$, weak screening can contribute to increase $\zetas$ sufficiently to cause observable hysteresis. In the strong-screening regime, both $\zetao$ and $\zetas$ increasingly become sensitive to local segmental friction as correlation blobs shrink towards the Kuhn segment with increasing concentration. 

The observations reported here have implications for understanding viscoelastic behaviour of polymer solutions well beyond the somewhat esoteric phenomenon of coil--stretch hysteresis. Intermolecular interactions are ignored in classical constitutive models used for modeling  of viscoelastic flows of nominally dilute solutions. The assumption is that dilute solutions stay dilute.  Any polymer solution with a concentration such that $c L^3 > 1$ can be expected to experience significant intermolecular interaction. Analytical arguments suggest that in general the contribution of the dissolved polymer to macroscopic stresses becomes significant just when intermolecular hydrodynamic interactions become important \citep{ottbk, Prabhakar2016}. Therefore, fully predicting the influence of polymer  concentration in applications  may require constitutive models that explicitly account for intermolecular hydrodynamic interactions. The blob model for average chain friction can be used to construct a microstructural consititutive model for viscoelastic stresses in unentangled solutions. Such a model has been demonstrated recently \citep{Prabhakar2016}  to predict the complex concentration dependence of the dynamics of capillary thinning of liquid bridges \citep{clasenetal}. Another question where such a model may be useful concerns the concentration at which polymer-induced turbulent drag reduction sets in. It has been shown that the dynamical slowdown associated with coil--stretch hysteresis can significantly influence chain dynamics in turbulent flows \citep{celani, Ahmad2016}.  The non-trivial effect  of polymer stretching that we have demonstrated here underlines the need for a more detailed examination of hydrodynamic screening in stretched chains.  Multi-chain simulations can be used to test the prediction from blob arguments that transverse chain overlaps do not contribute strongly to screening until correlation blobs become smaller than tension blobs. The insights from such studies could help refine blob-based constitutive models for unentangled polymer solutions.

\acknowledgments{This work was supported by a CPU-time grant on the National Computational Infrastructure at the Australian National University, Canberra.}


\end{document}